\input{aipcheck}

\documentclass[
    ,final            
  ]
  {aipproc}

\layoutstyle{8x11double}

\newcommand{\be}{\begin{equation}}
\newcommand{\ee}{\end{equation}}

\newcommand{\eq}[1]{Eq. (\ref{#1})}

 \newcommand{\eqa}{\begin{eqnarray}}
\newcommand{\eeq}{\end{eqnarray}}

\begin{document}

\title{Whistler Wave Turbulence  in Solar Wind Plasma}



\author{Dastgeer Shaikh\footnote{\tt Email:dastgeer.shaikh@uah.edu}}{
  address={Department of Physics and Center for Space Plasma and Aeronomy Research (CSPAR), \\
The University of Alabama in Huntsville, Huntsville, AL 35899, USA}
}

\author{G. P. Zank}{}

\begin{abstract}
Whistler waves are present in solar wind plasma. These waves possess
characteristic turbulent fluctuations that are characterized typically
by the frequency and length scales that are respectively bigger than
ion gyro frequency and smaller than ion gyro radius. The electron
inertial length is an intrinsic length scale in whistler wave
turbulence that distinguishably divides the high frequency solar wind
turbulent spectra into scales smaller and bigger than the electron
inertial length.  We present nonlinear three dimensional, time
dependent, fluid simulations of whistler wave turbulence to
investigate their role in solar wind plasma. Our simulations find that
the dispersive whistler modes evolve entirely differently in the two
regimes. While the dispersive whistler wave effects are stronger in
the large scale regime, they do not influence the spectral cascades
which are describable by a Kolmogorov-like $k^{-7/3}$ spectrum. By
contrast, the small scale turbulent fluctuations exhibit a
Navier-Stokes like evolution where characteristic turbulent eddies
exhibit a typical $k^{-5/3}$ hydrodynamic turbulent spectrum. By
virtue of equipartition between the wave velocity and magnetic fields,
we quantify the role of whistler waves in the solar wind plasma
fluctuations.
\end{abstract}

\maketitle


\section{1. Introduction}

The solar wind is an excellent in-situ laboratory for investigating
nonlinear and turbulent processes in a magnetized plasma fluid since
it comprises a multitude of spatial and temporal length-scales
associated with an admixture of waves, fluctuations, structures and
nonlinear turbulent interactions.  The in-situ spacecraft measurements
\cite{Brown,Goldstein1995} reveal that the solar
wind fluctuations, extending over several orders of magnitude in
frequency and wavenumber, describe the power spectral density (PSD)
spectrum that can be divided into three distinct
regions \cite{Goldstein1995,leamon}. The frequencies, for instance,
smaller than $10^5$ Hz lead to a PSD that has a spectral slope of -1
. This follows the region that extends from $10^5$ Hz to or less than
ion/proton gyrofrequency where the spectral slope exhibits an index of
-3/2 or -5/3.  Smaller than ion gyro radius ($k\rho_i
\gg 1$) and temporal scales bigger than ion cyclotron frequency
$\omega> \omega_{ci}=eB_0/m_ec$, (where $k, \rho_i, \omega_{ci}, e,
B_0, m_e, c$ are respectively characteristic mode, ion gyroradius, ion
cyclotron frequency, electronic charge, mean magnetic field, mass of
electron and speed of light) exhibit a spectral break where the
inertial range slope of the solar wind turbulent fluctuations varies
between -2 and -5 \cite{Goldstein1995,leamon}.  The onset of the
second or the kinetic Alfven inertial range is still elusive to our
understanding of the solar wind turbulence and many other nonlinear
interactions. Specifically, the mechanism leading to the spectral
break has been thought to be either mediated by the kinetic Alfven
waves (KAWs) \cite{hasegawa}, or by electromagnetic whistler
fluctuations \cite{gary,yoon}, or by a class of fluctuations that can
be dealt within the framework of the HMHD plasma
model \cite{alex,shaikh09}. Stawicki et al \cite{Stawicki} argue that
Alfv\'en fluctuations are suppressed by proton cyclotron damping at
intermediate wavenumbers so the observed power spectra are likely to
consist of weakly damped magnetosonic and/or whistler waves which are
dispersive unlike Alfv\'en waves. Moreover, turbulent fluctuations
corresponding to the high frequency and $k\rho_i \gg 1$ regime lead to
a decoupling of electron motion from that of ion such that the latter
becomes unmagnetized and can be treated as an immobile neutralizing
background fluid.  While whistler waves typically survive in the
higher frequency (and the corresponding smaller length scales) part of
the solar wind plasma spectrum, their role in influencing the inertial
range turbulent spectral cascades is still debated
\cite{biskamp,dastgeer03,dastgeer05,dastgeer08,dastgeer08a}.

In this paper, we focus on understanding the nonlinear turbulent
cascades mediated by whistler waves in a fully three dimensional
geometry. Our objective is to investigate the role of whistlers in
establishing the turbulent equipartition amongst the modes that are
responsible for the nonlinear mode coupling interactions which
critically determine the inertial range power spectra.

\section{2. Whistler wave model}
Whistler modes are excited in the solar wind plasma when the
characteristic plasma fluctuations propagate along a mean or background
magnetic field with frequency $\omega>\omega_{ci}$ and the length
scales are $c/\omega_{pi} < \ell < c/\omega_{pe}$, where $\omega_{pi},
\omega_{pe}$ are the plasma ion and electron frequencies.  The
electron dynamics plays a critical role in determining the nonlinear
interactions while the  ions merely provide a stationary
neutralizing background against fast moving electrons and behave as
scattering centers. The whistler wave turbulence can be described by
the electron magnetohydrodynamics (EMHD) model of plasma \citep{model}.
The
three-dimensional equation of EMHD describing the evolution of the
magnetic field fluctuations in whistler wave,
\eqa
\label{emhd3}
\frac{\partial }{\partial t}({\bf B}-d_e^2 \nabla^2 {\bf B} ) + {\bf
V}_e\cdot \nabla ({\bf B}-d_e^2 \nabla^2 {\bf B} )- \\ \nonumber ({\bf B}-d_e^2
\nabla^2 {\bf B} ) \cdot \nabla{\bf V}_e 
 = \mu d_e^2 \nabla^2 {\bf B}.
\eeq 
The length scales in \eq{emhd3} are normalized by the electron skin
depth $d_e = c/\omega_{pe}$ i.e. the electron inertial length scale,
the magnetic field by a typical amplitude ${B}_0$, and time by
the corresponding electron gyro-frequency.  In \eq{emhd3}, the
diffusion operator on the right hand side is raised to $2n$. Here $n$
is an integer and can take $n=1,2,3, \cdots$.  The case $n=1$ stands
for normal diffusion, while $n=2,3, \cdots$ corresponds to hyper- and
other higher order diffusion terms.

The linearization of \eq{emhd3} about a constant magnetic
field $B_0$ yields the dispersion relation for the whistlers, the
normal mode of oscillation in the EMHD frequency regime, and is given
by
\[
\omega_k = \omega_{c_0}\frac{d_e^2 k_yk}{1+d_e^2k^2},
\]
where $ \omega_{c_0}=eB_0/mc$ and $k^2=k_x^2+k_y^2$.  From \eq{emhd3},
it appears that there exists an intrinsic length scale $d_e$, the
electron inertial skin depth, which divides the entire spectrum into
two regions; namely short scale ($kd_e>1$) and long scale ($kd_e<1$)
regimes.  In the regime $kd_e<1$, the linear frequency of whistlers is
$\omega_k \sim k_y k$ and the waves are dispersive.  Conversely,
dispersion is weak in the other regime $kd_e>1$ since $\omega_k \sim
k_y/ k$ and hence the whistler wave packets interact more like the
eddies of hydrodynamical fluids.

\begin{figure}
\label{fig1}
  \includegraphics[height=.25\textheight]{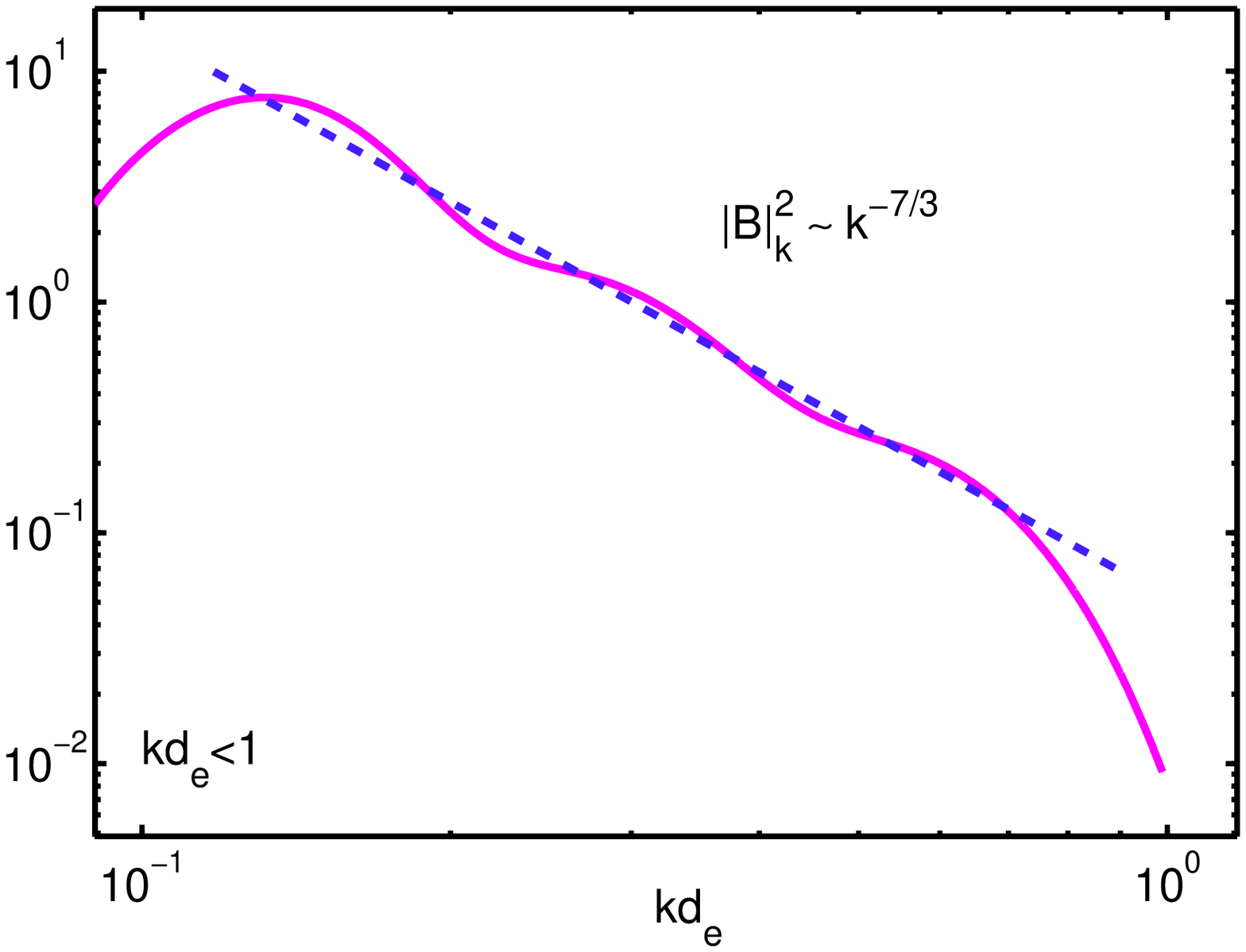}
\includegraphics[height=.25\textheight]{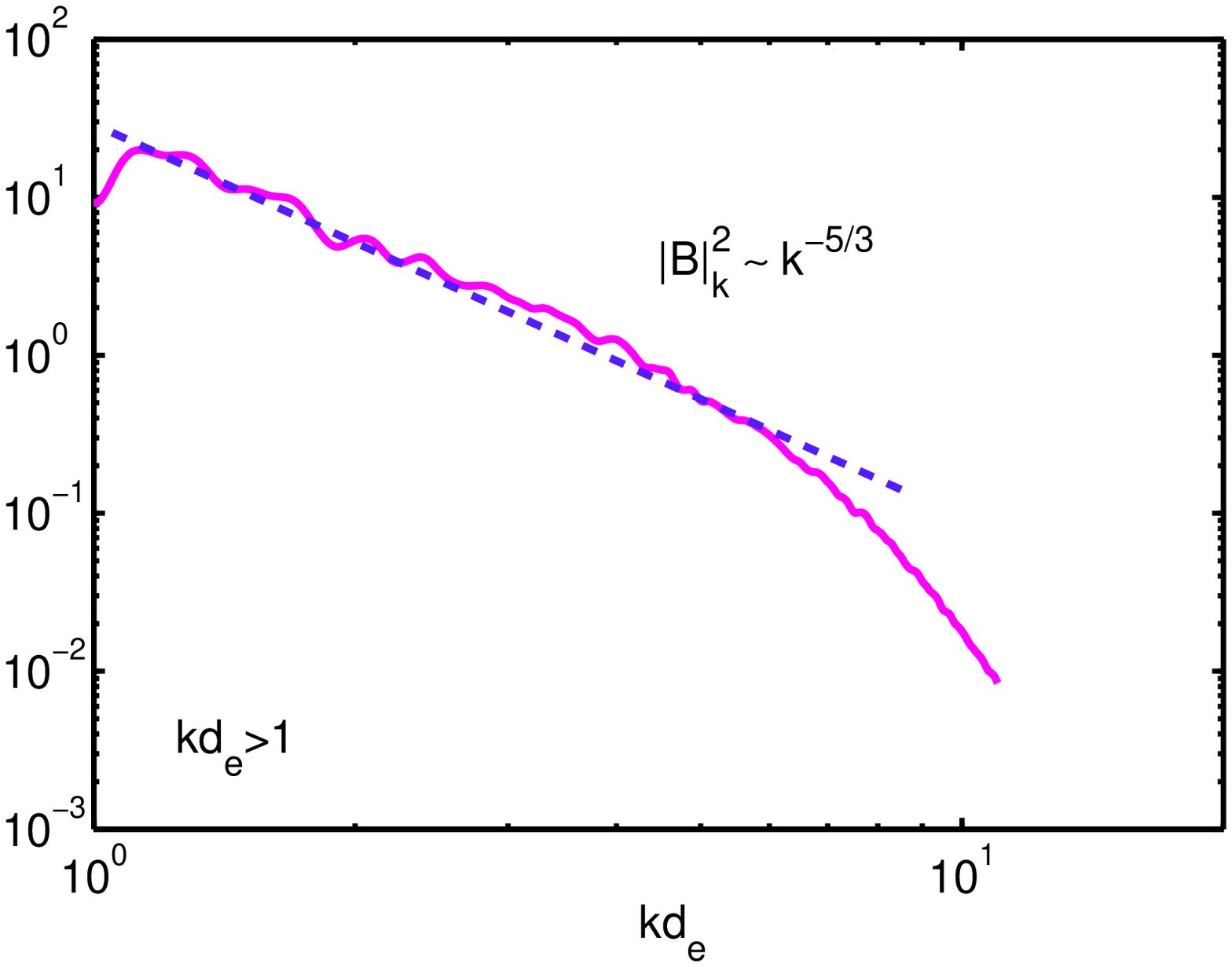}
  \caption{(LEFT) 3D simulation of whistler wave turbulence in the $kd_e<1$
  regime exhibits a Kolmogorov-like inertial range power spectrum
  close to $k^{-7/3}$. The simulation parameters are: Box size is $L_x
  \times L_y \times L_z=2\pi \times 2\pi \times 2\pi$, numerical
  resolution is $N_x \times N_y \times N_z = 200 \times 200 \times
  200$, electron skin depth is $d_e=0.015$, magnitude of constant
  magnetic field is $B_0=0.5$. (RIGHT) the $kd_e>1$
  regime depicts a Kolmogorov-like $k^{-5/3}$ spectrum.  $d_e=0.15$.}
\end{figure}

\section{3. Simulations}
Turbulent interactions mediated by the coupling of whistler waves and
inertial range fluctuations are studied in three dimensions (3D) based
on a nonlinear 3D whistler wave turbulence code that we have developed
at Center for Space Plasma and Aeronomic Research (CSPAR), the
University of Alabama in Huntsville (UAH). Our code numerically
integrates \eq{emhd3}.  The spatial descritization employs a
pseudospectral algorithm \cite{scheme,dastgeer03,dastgeer08} based on
a Fourier harmonic expansion of the bases for physical variables
(i.e. the magnetic field, velocity), whereas the temporal integration
uses a Runge Kutta (RK) 4th order method.  The boundary conditions are
periodic along the $x,y$ and $z$ directions in the local rectangular
region of the solar wind plasma.

Electron whistler fluid fluctuations, in the presence of a constant
background magnetic field, evolve by virtue of nonlinear interactions
in which larger eddies transfer their energy to smaller ones through a
forward cascade. According to \cite{kol}, the cascades of spectral
energy occur purely amongst the neighboring Fourier modes (i.e. local
interaction) until the energy in the smallest turbulent eddies is
finally dissipated gradually due to the finite dissipation. This leads
to a damping of small scale motions.  By contrast, the large-scales
and the inertial range turbulent fluctuations remain unaffected by
direct dissipation of the smaller scales. Since there is no mechanism
that drives turbulence at the larger scales in our model, the
large-scale energy simply migrates towards the smaller scales by
virtue of nonlinear cascades in the inertial range and is dissipated
at the smallest turbulent length-scales.  The spectral transfer of
turbulent energy in the neighboring Fourier modes in whistler wave
turbulence follows a Kolmogorov phenomenology \citep{kol,iros,krai}
that leads to Kolmogorov-like energy spectra. We find from our 3D
simulations that whistler wave turbulence in the $kd_e<1$ and $kd_e>1$
regimes exhibits respectively $k^{-7/3}$ and $k^{-5/3}$ (see Fig 1)
spectra. The inertial range turbulent spectra obtained from our 3D
simulations are further consistent with 2D work
\cite{biskamp,dastgeer00a,dastgeer00b}.  Interestingly,  the wave effects
dominate in the large scale, i.e. $kd_e<1$, regime where the inertial
range turbulent spectrum depictes a Kolmogorov-like $k^{-7/3}$
spectrum. On the other hand, turbulent fluctuations in the smaller
scale ($kd_e>1$) regime behave like non magnetic eddies of
hydrodynamic fluid and yield a $k^{-5/3}$ spectrum.  The wave effect
is weak, or negligibly small, in the latter. Hence the nonlinear
cascades are determined essentially by the hydrodynamic like
interactions. The observed whistler wave turbulence spectra in the
$kd_e<1$ and $kd_e>1$ regimes (Figs 1) can be followed from the
Kolmogorov-like arguments \citep{kol, iros, krai} that describe the
inertial range spectral cascades. We elaborate on these arguments to
explain our simulation results of Fig. (1) in the following section.

\section{4. Whistler wave spectra}
The exact spectral indices corresponding to the whistler wave
turbulent spectra, described by the ideal electron magnetohydrodynamic
invariant, can be understood from the Kolmogorov's dimensional
arguments \cite{kol, iros, krai}.

In the underlying whistler wave model of magnetized plasma turbulence,
the inertial range eddy velocity is characterized typically by $v_e
\sim \nabla \times{\bf B}$. Thus the typical velocity of the magnetic
field eddy $B_{\ell}$ with a scale size $\ell$ can be represented by
$v_e \simeq B_{\ell}/\ell$. The eddy turn-over time is then given by
\[\tau \sim \frac{\ell}{v_e} \sim \frac{\ell^2}{B_{\ell}}.\]
This is the time scale that predominantly leads to the nonlinear
spectral transfer of energy in fully developed whistler wave
turbulence.

In the regime where characteristic length scales are bigger than the
electron skin depth ($kd_e < 1$), the inertial range whistler
turbulent energy is dominated by the large scale fluctuations. The
total energy corresponding to the turbulent fluctuations in this
regime is then given as,
\[ E \sim |{\bf B}|^2 \sim B_{\ell}^2 \sim v_e^2 \ell^2. \]
The $B_{\ell}$ represent magnetic field associated with the magnetic
field eddy of length $\ell$. The second similarity follows from the
assumption of an equipartition of energy in the magnetic and velocity
field components of whistler waves. The process of equipartition
origintes from the correlation between the velocity and magnetic field
fluctuations $v_e \sim {\bf k} \times{\bf B}$, where ${\bf
k}=k_x\hat{x}+k_y \hat{y}+ k_z \hat{z}$ is a three dimensional wave
vector. The latter is further consistent with the electron flow speed,
in combination with the wave perturbed magnetic field, that is used to
derive the dynamical equation of whistler wave turbulence,
i.e. \eq{emhd3}.  This velocity-magnetic field correlation essentially
produces the velocity field fluctuations that are normal to the
magnetic field in a whistler wave packet. Consequently, the energy
associated with the velocity and magnetic field for each
characteristic turbulent mode evolves toward a relationship that
satisfies $v_e^2
\simeq k^2 B^2$. To quantify our arguments, we follow the
evolution of turbulent equipartion in our simulations by computing the
following quantity, 
\be E_{equi}(t) \simeq \sum_k (|v_e(k,t)|^2 -
k^2|B(k,t)|^2),
\label{equi}
\ee
 Interestingly, our 3D
simulations, describing the equipartition between the velocity and
magnetic field fluctuations, are consistent with the 2D counterpart.
It thus appears that the turbulent equipartition is a robust feature
of whistler waves that is preserved in both 2D and 3D nonlinear mode
coupling interactions.  The spectral cascades of inertial range
turbulent energy is nonetheless determined by the energy cascade per
unit nonlinear time as follows,
\[ \varepsilon \simeq \frac{E}{\tau} \simeq \frac{B_{\ell}^3}{\ell^2}. \]
On assuming that the spectral energy cascade is local in the
wavenumber space \cite{iros,kol,krai}, the energy spectrum per unit
mode yields $ E_k \simeq \varepsilon^{\alpha}k^{\beta}$. On substituting
the energy and energy dissipation rates and equating the powers of
$B_{\ell}$ and $\ell$, we obtain $\alpha = 2/3$ and $\beta = -7/3$. This,
in the $kd_e < 1$ regime, leads to the following expression for the
energy spectrum $ E_k \simeq \varepsilon^{2/3}k^{-7/3}$.

\begin{figure}
\label{fig2}
  \includegraphics[height=.3\textheight]{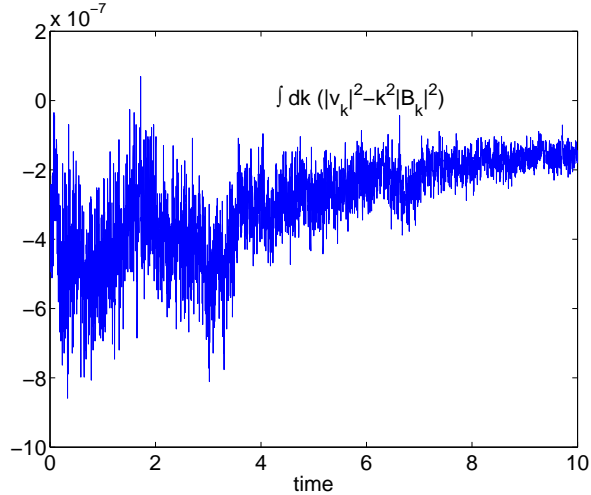}
  \caption{Turbulent equipartion between the velocity and magnetic
  fields is observed in our 3D simulations. When the
  characteristic turbulent modes evolve towards equipartition, the
  relationship $|v_{e}({\bf k},t)|^2 \simeq k^2|B({\bf k},t)|^2$ is
  obeyed. Consequently, $E_{equi} \rightarrow 10^{-7}$.}
\end{figure}

\section{5. Conclusions}
Three dimensional simulations of turbulent cascades in solar wind
plasma are carried out to quantify the role of whistler waves
corresponding to the inertial range fluctuations that possess
characteristic frequency bigger than the ion gyro frequency ($\omega >
\omega_{ci}$) and length scales smaller than the ion gyro radius
($k\rho_i>1$). In this regime, the solar wind plasma fluctuations
comprise of unmagnetized ions, hence the entire dynamics is governed
by the electron fluid motions. The rotational magnetic field
fluctuations in the presence of a background magnetic field lead to
propagation of dispersive whistler waves in which the wave magnetic
and velocity fields are strongly correlated through the equipartition
($v_e^2 \simeq k^2 B^2$). The latter is employed in our simulations to
quantify the role of whistler waves that are ubiquitously present in
the inertial range in the high frequency ($\omega > \omega_{ci}$)
solar wind plasma. Interestingly we find that despite strong wave
activity in the inertial range, whistler waves do not influence the
inertial range turbulent spectra. Consequently, the turbulent
fluctuations in the inertial range are described by Kolmogorov-like
phenomenology \cite{kol}. 

It is to be noted that as long as the cascade of energy is concerned,
kinetic \cite{gary,saito} and
fluid \cite{dastgeer03,dastgeer05,dastgeer08,dastgeer08a} like
processes lead to a similar power law (i.e. $E_k \sim k^{-5/3}$) in
the $kd_e>1$ regime. This is because the energy cascade is determined
entirely by the {\em convective time scales} that are associated with
the nonlinear term ${\bf V}_e \cdot \nabla \nabla^2 {\bf B}$ in the
electron fluid momentum equation. The breakdown of fluid-like behavior
occurs for the characteristic scales $kd_e \gg 1, k\rho_e \sim 1$,
where $\rho_e$ is electron gyro radius.  The major difference in the
fluid and kinetic models, however, arises from wave-particle
resonances (or wave particle interactions) which are a fully kinetic
effect and it is beyond the capability of the fluid theory. Since the
energy spectra are not critically dependent on the wave-particle
resonances, our fluid model yields spectral laws similar to the
kinetic model.

To conclude, consistent with the Kolmogorov-like dimensional
argument \cite{kol}, we find that turbulent spectra in the $kd_e<1$
and $kd_e>1$ regimes are described respectively by $k^{-7/3}$ and
$k^{-5/3}$. Our results are important particularly in understanding
turbulent cascade corresponding to the high frequency ($\omega
> \omega_{ci}$) solar wind plasma where characteristic fluctuations
are comparable to the electron inertial skin depth.


The support of NASA(NNG-05GH38) and NSF (ATM-0317509) grants is  acknowledged.





\IfFileExists{\jobname.bbl}{}
 {\typeout{}
  \typeout{******************************************}
  \typeout{** Please run "bibtex \jobname" to optain}
  \typeout{** the bibliography and then re-run LaTeX}
  \typeout{** twice to fix the references!}
  \typeout{******************************************}
  \typeout{}
 }


\end{document}